\documentclass{nature_mod}
\usepackage{graphicx}
\usepackage{graphics}
\usepackage{ulem}

\usepackage{amsmath,amssymb}

\newcommand\kms{{\rm\,km\,s^{-1}}}
\newcommand\msun{\rm\,M_\odot}
\newcommand\lsun{\rm\,L_\odot}

\title{A solar-type star polluted by calcium-rich supernova ejecta inside the supernova remnant RCW\,86}

\author{Vasilii V.\ Gvaramadze$^{1,2,3,*}$, Norbert Langer$^{4}$, Luca Fossati$^5$, Douglas C.-J.\ Bock$^6$, Norberto Castro$^{4,7}$,
Iskren Y.\ Georgiev$^8$, Jochen Greiner$^9$, Simon Johnston$^6$,
Arne Rau$^9$ \& Thomas M.\ Tauris$^{10,4}$}

\begin{document}

\maketitle

\begin{affiliations}
 \item Sternberg Astronomical Institute, Lomonosov Moscow State University, Universitetskij Pr.~13, Moscow 119992, Russia
 \item Space Research Institute, Russian Academy of Sciences, Profsoyuznaya 84/32, 117997 Moscow, Russia
 \item Isaac Newton Institute of Chile, Moscow Branch, Universitetskij Pr.\ 13, Moscow 119992, Russia
 \item Argelander-Institut f\"ur Astronomie, Auf dem H\"{u}gel 71, 53121 Bonn, Germany
 \item Space Research Institute, Austrian Academy of Sciences, Schmiedlstrasse 6, 8042 Graz, Austria
 \item CSIRO Astronomy and Space Science, Australia Telescope National Facility, PO Box 76, Epping, NSW 1710, Australia
 \item Department of Astronomy, University of Michigan, 1085 S. University Avenue, Ann Arbor, MI 48109, USA
 \item Max-Planck Institut f\"{u}r Astronomie, K\"{o}nigstuhl 17, 69117 Heidelberg, Germany
 \item Max-Planck-Institut f\"{u}r extraterrestrische Physik, Giessenbachstr. 1, 85748 Garching, Germany
 \item Max-Planck-Institut f\"{u}r Radioastronomie, Auf dem H\"{u}gel 69, 53121 Bonn, Germany
\end{affiliations}

\begin{abstract}
When a massive star in a binary system explodes as a supernova its companion star
may be polluted with heavy elements from the supernova ejecta. Such a pollution had
been detected in a handful of post-supernova binaries\cite{Gon11}, but none of them 
is associated with a supernova remnant. We report the discovery of a solar-type star 
in a close, eccentric binary system with a neutron star within the young Galactic 
supernova remnant RCW\,86. Our discovery implies that the supernova progenitor was 
a moving star, which exploded near the edge of its wind bubble and lost most of its 
initial mass due to common-envelope evolution shortly before core collapse. We find 
that the solar-type star is strongly polluted with calcium and other elements, which 
places the explosion within the class of calcium-rich supernovae -- faint and fast 
transients\cite{Fil03, Kas12}, whose origin is strongly debated\cite{Kaw10, Wal11}, 
and provides the first observational evidence that supernovae of this type can arise 
from core-collapse explosions\cite{Kaw10, Mor10}.
\end{abstract}

Recently, it was recognized that a large fraction of massive stars
reside in binary systems\cite{San12, Chi12} and that the evolution
of the majority of massive stars is strongly affected by binary 
interaction, through 
mass transfer, common envelope evolution or merger\cite{San12}.
This suggests that most Type\,Ib/c supernovae (SNe), which do not
show hydrogen lines in their spectra, stem from progenitors which
are stripped of their hydrogen envelopes by a companion star
\cite{Pod92,Lan12}. The low ejecta masses typical of such
SNe\cite{Lym16a} imply that a significant fraction of the post-SN
binaries remain bound. Only very few of such binaries are
known\cite{Bha09, Hei13} to be associated with supernova remnants
(SNRs) because of the short lifetime ($\sim$$10^5$ yr) of the
SNRs. In this Letter, we report the discovery of a post-SN binary
system within the young (few thousand years\cite{Dic01}) SNR
RCW\,86.

The pyriform appearance of RCW\,86 (Fig.\,\ref{fig:rcw86}; see
also fig.\,6 in ref.\cite{Smi97}) can be explained as the result
of a SN explosion near the edge of a bubble blown by the wind of a
moving massive star\cite{Gva03, Mey15} (Supplementary Information
section\,1). This interpretation implies that the SN exploded near
the centre of the hemispherical optical nebula in the south-west
of RCW\,86 (see Fig.\,\ref{fig:rcw86}) and that the stellar
remnant should still be there. Motivated by these arguments, we
looked for a possible compact X-ray source using archival {\it
Chandra} data and discovered\cite{Gva03} two sources in the
expected position of the SN progenitor (Fig.\,\ref{fig:rcw86}).
One of them, [GV2003]\,S, has a clear optical counterpart with
$V$=14.4 mag and its X-ray spectrum implies that this source is a
foreground late-type active star. For the second source,
[GV2003]\,N, we did not find any optical counterpart in the
Digital Sky Survey\,II to a limiting red band magnitude of
$\approx$21, while its X-ray spectrum suggests that this source
could be a young pulsar\cite{Gva03}. Our deep follow-up
observation with the Parkes radio telescope in 2002, however,
failed to detect any radio emission from [GV2003]\,N, giving an
upper limit on the flux of 35 $\mu$Jy at 1420 MHz (see Methods).
This non-detection may be a consequence of beaming or it could
indicate that [GV2003]\,N may not be an active radio pulsar.

If [GV2003]\,N was a NS its emission in the visual was expected to
be fainter than $V$$\approx$28\,mag. We therefore obtained a
$V$-band image of the field around this source with the FORS2
instrument on the ESO Very Large Telescope (VLT) in 2010. The
FORS2 image, however, revealed a stellar-like object with
$V$=20.69$\pm$0.02 mag just at the position of [GV2003]\,N
(Fig.\,\ref{fig:rcw86}; Methods). To further constrain the nature
of [GV2003]\,N, we obtained its $g'r'i'z'JHK_{\rm s}$ photometry
with the 7-channel optical/near-infrared imager GROND in 2013
(Methods). With that, we fitted the spectral energy distribution
(SED) of [GV2003]\,N and derived a temperature of $\approx$5200\,K
and a colour excess of $E(B-V)$$\approx$0.9 mag (Methods; Extended
Data Fig.\,1). These results exclude the possibility that
[GV2003]\,N is an AGN (as it was proposed in ref.\cite{Mig12}) and
strongly suggest that the optical emission originates from a
G-type star at a distance comparable to that of RCW\,86 of
2.3$\pm$0.2 kpc\cite{Sol03}. Since the X-ray luminosity of
[GV2003]\,N of $\sim$$10^{32} \, {\rm erg} \, {\rm s}^{-1}$
(ref.\cite{Gva03, Mig12}) is far too high for a G
star\cite{Gud04}, we arrived at the possibility that we are
dealing with a G star orbiting the NS.

\begin{figure}
\caption{{\bf Portion of the VLT/FORS2 spectrum of the optical
counterpart of [GV2003]\,N.} The observed spectrum (dots connected
by a solid line) is compared with synthetic spectra calculated
from the derived atmospheric parameters for the solar (red dashed
line) and the final estimated abundances (blue solid line). The
major components of each blend are labelled. One can see that the
solar abundances are not a good match to the observed spectrum and
that Ca and Fe are overabundant. \label{fig:spec} }
\begin{center}
\includegraphics[width=14cm,clip]{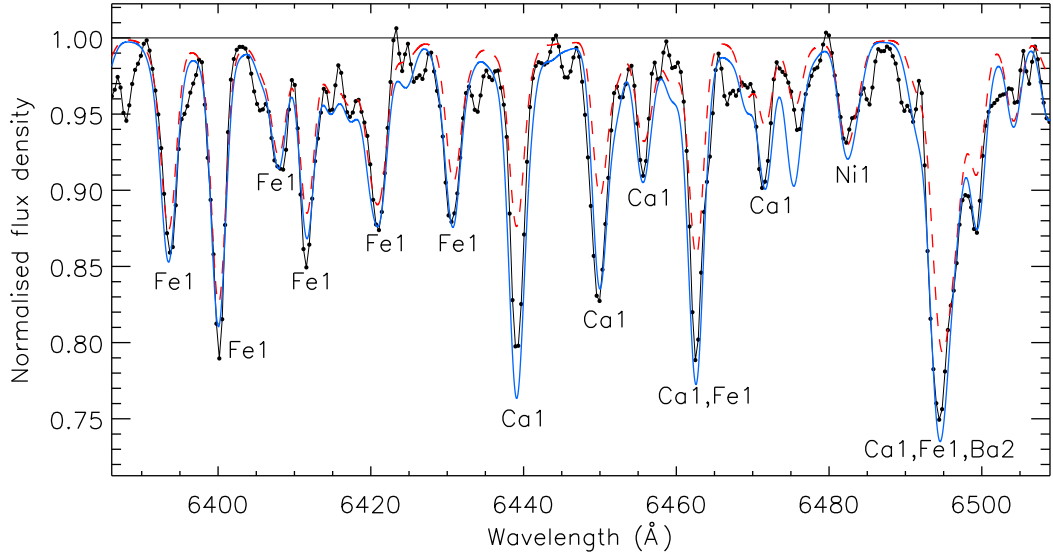}
\end{center}
\end{figure}

Consequently, we searched for radial velocity (RV) variability and
traces of the SN ejecta in the spectrum of the optical counterpart
of [GV2003]\,N. We obtained four spectra with the VLT/FORS2 in
2015. We found clear RV variations (Table\,\ref{tab:RV}),
indicative of an eccentric binary with a period of about a month
or less (Methods). The spectrum of [GV2003]\,N
(Fig.\,\ref{fig:spec}) clearly resembles that of a solar-type
star, confirming the results from the SED analysis.

From the analysis of the Fe\,{\sc i} lines (Methods), we derived a
stellar effective temperature of $T_{\rm eff}=5100$$\pm$200\,K.
Using $V$=20.69 mag, $E(B-V)$=0.9 mag and assuming the ratio of
total to selective extinction of $R_V$=3.1, we estimated a visual
extinction and absolute magnitude of $A_V$=2.79 mag and $M_V$=6.09
mag, respectively. With a bolometric correction of $-$0.29
mag\cite{Pec13}, we obtained a bolometric luminosity of $\log
(L/\lsun)=-$0.42$\pm$0.08, taking into account the uncertainty on
the distance. Using the Pisa stellar models\cite{Tog11}, we
estimated the mass of the optical star as $\approx 0.9 \, \msun$,
which for the given $T_{\rm eff}$ and $L$ corresponds to a surface
gravity of 4.6$\pm$0.1. The fact that the binary system remained
bound after the SN explosion, and assuming the standard mass of a
NS of $1.4 \, \msun$, implies a mass of the SN ejecta of less than
$2.3 \, \msun$ --- provided that the SN explosion was
symmetric\cite{Hil83} --- and a mass of the exploding star below
$3.7 \, \msun$. This in turn implies that the SN explosion was of
Type\,Ib and that the initial mass of the primary star was smaller
than $13 \, \msun$\cite{Tut73}.

From the FORS2 spectra we derived the abundances of Si, Ca, Ti, V,
Cr, Mn, Fe, Co, Ni, and Ba (Methods). Fig.\,\ref{fig:abund} shows
that many elements are enhanced by a factor of about 3 with
respect to the solar abundances\cite{Asp09}, with the silicon and
iron being less than doubled. Calcium is particularly
overabundant, by a factor of $\approx$6, which, to our knowledge,
makes [GV2003]\,N the most Ca-rich star known to date.

The SN ejecta captured by the G star are mixed with the material
of its convective envelope. Adopting an envelope mass of $0.2 \,
\msun$\cite{Dan94}, we computed the mass accreted from the SN
ejecta for each element with measured abundances (Extended Data
Table\,1), which sums up to $3.3\times10^{-4} \, \msun$. Since the
accreted mass of iron of $\approx$$1.5\times10^{-4} \, \msun$ was
ejected in the form of $^{56}$Ni, an accreted fraction of the SN
ejecta of about 1\% would lead to a total mass of radioactive
nickel produced by the SN of $0.015 \, \msun$. These numbers
appear consistent, since the explosion energy of SNe of $\sim$$10
\, \msun$ stars is thought to be low, with the consequence of a
relatively low nickel mass and moderate ejecta
velocities\cite{Suk16}, allowing for a rather high accreted
fraction of the SN ejecta (Supplementary Information section\,4).

The large amount of Ca in [GV2003]\,N, together with the rather
modest enhancement of Si suggest that the SN, which produced
RCW\,86, was not a typical core-collapse event (Supplementary
Information section 2). Instead, our data suggest that it belonged
to the class of Ca-rich SNe --- the fast and faint transients,
which show strong Ca lines in their spectra and are believed to be
intrinsically calcium-rich\cite{Per10} (Supplementary Information
section\,3). Whereas observations indicate that many of the
Ca-rich SNe are produced by long-lived low-mass stars, our
findings imply that the lowest mass stars capable of producing NSs
can produce similar explosion, if the progenitor is stripped of
its envelope by a companion star. Indeed, from theoretical
considerations, in both cases the explosion is produced in a
helium layer with an inner radius comparable to that of a white
dwarf, and an explosion kinetic energy of less than $10^{51}$\,erg
(Supplementary Information section\,3).

The short orbital period of [GV2003]\,N implies that this binary
system will evolve into a low-mass X-ray binary (LMXB) within its
nuclear time scale ($\sim$$10^{10}$\,yr), providing the first
definite example of a pre-LMXB located within a SNR. Before that,
thermohaline mixing is expected to dilute the accreted SN material
in the bulk of the mass of [GV2003]\,N on a time scale of
$\sim$$10^7$\,yr. Thereby, the enhanced abundances of the measured
elements will be reduced by factor of several, making them
comparable to those found in the majority of the few main sequence
components of LMXBs for which pollution has been detected
previously\cite{Gon11}.

\section*{Bibliography}
\vspace{1cm}

\bibliographystyle{naturemag}


\begin{addendum}
 \item
 Based on observations collected at the European Southern Observatory, Chile, under
 programmes 095.D-0061 and 385.D-0198(A). V.V.G. is grateful to M.G.~Revnivtsev for useful
 discussions, and acknowledges support from the Russian Science Foundation grant 14-12-01096.
 \item[Author Contributions]
V.V.G. and N.L. led the project and the manuscript writing.
V.V.G., N.L., L.F. and D.C.-J.B. wrote the telescope proposals.
L.F. reduced the VLT/FORS2 spectra and performed the spectral
analysis. S.J. and D.C.-J.B performed and analyzed the radio
observations. I.Y.G. performed the PSF photometry. J.G. and A.R.
performed the GROND observations and the SED fitting. N.C.
performed part of the absolute wavelength calibration of the
VLT/FORS2 spectra and worked on the removal of the spatially
variable H$\alpha$ emission.  T.M.T. performed the Monte Carlo
simulations of SN explosions in binary systems. Figures were
prepared by V.V.G., L.F., A.R and T.M.T. All authors contributed
to the interpretation of the data and commented on the manuscript.
  \item[Author Information] Reprints and permissions information is
 available at www.nature.com/reprints. The authors declare that they have no competing financial
interests. Correspondence and requests for materials should be
addressed
 to V.V.G. (vgvaram@mx.iki.rssi.ru) or N.L (nlanger@astro.uni-bonn.de).
  %
\end{addendum}

\newpage

\begin{figure}
\caption{{\bf SNR RCW\,86 and [GV2003]\,N.} From the upper left
clockwise: Molonglo Observatory Synthesis Telescope 843
MHz\cite{Whi96} image of RCW\,86, Digital Sky Survey II red band
image of an arc-like optical nebula in the south-west corner of
RCW\,86, and VLT/FORS2 and {\it Chandra} images of two point
sources, [GV2003]\,N and [GV2003]\,S, in the centre of the optical
arc (marked, respectively, by blue and magenta circles). White
spots in the VLT/FORS2 image are due to saturation effect. The
orientation of the images is the same. At the distance of RCW\,86
of 2.3 kpc, 10 arcmin and 5 arcsec correspond to $\approx$6.6 and
0.05 pc, respectively. \label{fig:rcw86} }
\begin{center}
\includegraphics[width=12cm]{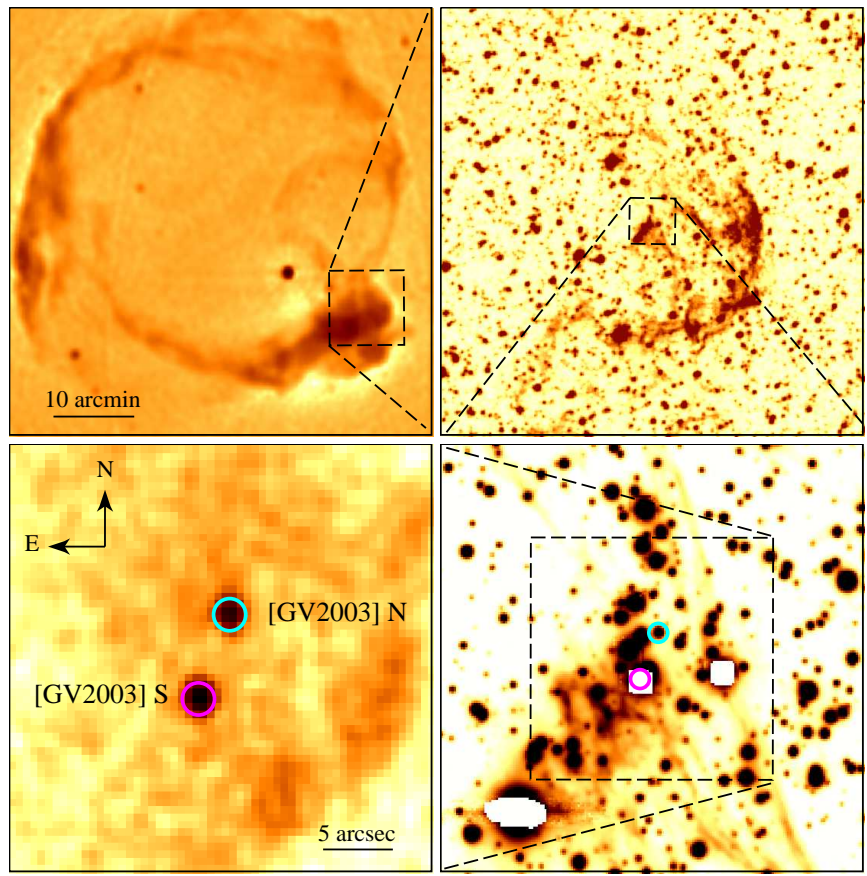}
\end{center}
\end{figure}

\newpage

\newpage

\begin{figure}
\caption{{\bf Element abundances of the optical counterpart of
[GV2003]\,N.} The dashed line corresponds to solar abundance
values. The smaller error bars indicate the statistical
uncertainties, while the larger ones show the maximum systematic
uncertainties. The real uncertainties lie in between. For the Li
abundance we give only an upper limit. Note a very high
overabundance of Ca. \label{fig:abund} }
\begin{center}
\includegraphics[width=14cm,clip]{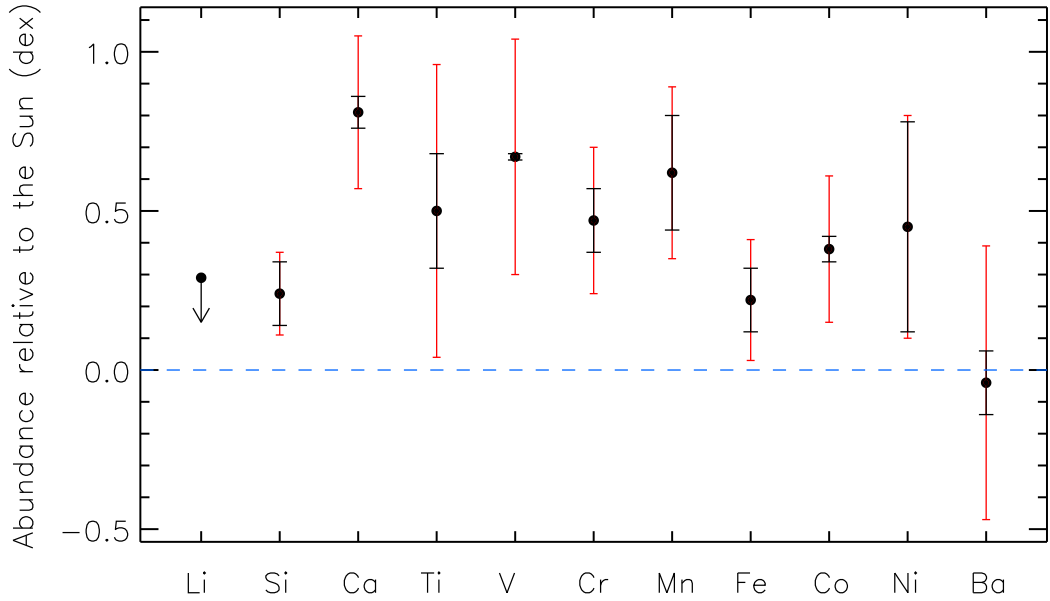}
\end{center}
\end{figure}

\newpage

\begin{methods}

\subsection{Radio observations of [GV2003]\,N.}
Observations were conducted in the L-band at the Parkes radio
telescope on 2002 October 13. The bandwidth was 256 MHz, with 512
spectral channels. We observed [GV2003]\,N for about 7.5 hours,
with a time resolution of 0.4 ms. The recorded data were
dedispersed for 512 independent dispersion measures (DMs),
uniformly spaced between zero and 275 pc cm$^{-3}$, beyond which
the smearing of the signal in the individual spectral channels
becomes greater than the sampling interval. Then, we searched for
harmonically related peaks by the standard harmonic folding
procedure, after performing a Fourier transform, for the time
series corresponding to each DM. Details of statistically
significant ``candidates'' were stored, and later used to produce
coherently folded profiles for assessing the quality. The
dedispersion of the data was performed with Taylor's tree
algorithm\cite{Tay74}, which assumes linearity of the dispersion
delay as a function of frequency. In our case, the frequency was
sufficiently high for the error due to this assumption to be
negligible ($\sim$1.2\%), and we did not need to apply further
correction to the band.

Ultimately, we did not detect any reliable (pulsed) radio signal
from [GV2003]\,N, although our search was very sensitive. In
general, detection sensitivity for pulsars is difficult to
quantify, mainly because of uncertainties introduced by unknown
duty cycle of pulsars, interstellar scattering, and the details of
the search procedure. The sensitivity of our observations was
estimated as follows. The telescope gain, $G$, of Parkes is
$\approx$$0.65 \, {\rm K} \, {\rm Jy}^{-1}$. The system
temperature, $T_{\rm sys}$, in the L-band is $\approx$32\,K, with
a sky contribution of about 6\,K. In general, one expects the
intrinsic duty cycle of long period pulsars to be $\sim$4\%.
Interstellar scatter broadening and dispersion smearing will
increase this width for high DM pulsars (these smearings add in
quadrature with the pulse width). The minimum detectable flux can
be written as $S_{\rm min}=(AT_{\rm sys}/G)[w/(N_{\rm pol}\beta
T)(P-w)]^{1/2}$, with $A$, $\beta$, $N_{\rm pol}$ and $T$ being
detection threshold in terms of r.m.s. of the noise, bandwidth,
number of independent polarization channels, and total observation
time, respectively. With a bandwidth of 256 MHz and observing time
of 7.46 hrs ($2^{26}$ samples at 0.4 mc intervals), this amounts
to 35 $\mu$Jy with a 10$\sigma$ limit. In comparison, this is
about 3 times more sensitive than the Parkes Multi-beam Pulsar
Survey\cite{Man01}.

\subsection{VLT/FORS2 imaging of [GV2003]\,N.}
%
[GV2003]\,N was observed with the FOcal Reducer and low dispersion
Spectrograph (FORS2) camera\cite{App98} at the ESO VLT in the $V$
band on 2010 April 10--12. With the FORS2 resolution of 0.25
arcsec per pixel, we obtained high quality images from a total of
45 exposures of 300\,s each in dark time, photometric conditions
with seeing of $\approx$0.6 arcsec and an average airmass of
$\approx$1.3.

The image reduction was performed with the software package {\sc
iraf}\cite{Tod93}. We created master bias and master normalized
sky flat field calibration images. Detector bad pixel map was
constructed from the ratio of low and high-count rate flats. These
calibration products were used to process each individual $V$-band
exposure. Single images were registered to a reference frame with
the smallest airmass. Geometric transformation solutions were
estimated with {\sc geomap} and {\sc geotran} procedures using
over a 1500 reference stars in common. The final average combined
(with {\sc ccdclip} algorithm) image was obtained from all images
brought to the same zero (sky) level as the reference using the
mode value of a 25 pixel $\times$ 25 pixel statistics region. The
large number of single exposures and the 9-point observing dither
pattern allowed us to create a high S/N final average combined
image, clean from remaining detector cosmetics (bad pixels and
rows/columns) and cosmic ray hits. The full-width at half maximum
(FWHM) of this image is 2.1 pixels, i.e. 0.53 arcsec.

We performed a point spread function (PSF) photometry on all
sources in the final combined image. For that we created a PSF
model using 168 isolated stars (no contaminating sources within a
radius of 3 FWHM). Best photometry was obtained from a second
order variable PSF model dependent on the position on the
detector, resulting in negligible ($<$1\% in flux) residuals. The
aperture and PSF photometry radius was chosen to be $\sim$3
FWHM$\approx$6 pixels, where the sky value was estimated locally
from an annulus with a width of 5 pixels. Aperture correction to
the PSF magnitudes was estimated from comparison between the PSF
and aperture magnitudes of the isolated PSF stars used to build
the PSF model. We perform PSF fitting photometry to all sources
with detection threshold of 4$\sigma$ above the background.
Optical counterpart to [GV2003]\,N was readily detected (see
Fig.\,\ref{fig:rcw86}).

Photometric Stetson standard star fields of NGC\,2437 and E5 (with
53 and 75 stars, respectively) were taken at the beginning and end
of each night. These were used to obtain the photometric zeropoint
($V_{\rm ZP}$=28.1563$\pm$0.0013 mag) and airmass coefficients
($X_{\rm V}$=0.1996$\pm$0.0014 mag) to convert from instrumental
to standard magnitudes. The aperture corrected instrumental PSF
magnitudes were calibrated using these coefficients at an airmass
of 1.269. We found that the apparent $V$-band magnitude of
[GV2003]\,N is 20.688$\pm$0.024.

\subsection{GROND photometry and SED fitting.}
[GV2003]\,N was observed on 2013 April 8 simultaneously in 4
optical ($g^{\prime}, r^{\prime}, i^{\prime}, z^{\prime}$) and 3
near-infrared ($J$, $H$, $K_{\rm s}$) bands with the GROND
instrument\cite{Gre08} at the 2.2\,m MPG telescope at the ESO La
Silla Observatory (Chile). Five exposures were obtained with
combined integration of $\approx$41\,min in the optical bands and
$\approx$53\,min in the near-infrared bands. Observing conditions
were good with a medium seeing of 1 arcsec and an average airmass
of 1.2. The data reduction was performed using standard {\sc iraf}
tasks\cite{Tod93, Kru08}. The $g^{\prime}, r^{\prime}, i^{\prime},
z^{\prime}$ photometry was obtained using PSF fitting while due to
the undersampled PSF in the near-infrared, the $J, H, K_{\rm s}$
photometry was measured from apertures with the sizes
corresponding to the FWHM of field stars. Calibration of the
optical photometry was obtained using an observation of the same
field obtained in a different night under photometric conditions
and calibrated against an observation of an SDSS
field\cite{Aih11}. Photometric calibration of the near-infrared
bands was achieved against selected 2MASS\cite{Skr06} stars in the
field of the target. The resulting AB magnitudes are:
$r^{\prime}$=19.60$\pm$0.05 mag, $i^{\prime}$=18.77$\pm$0.08 mag,
$z^{\prime}$=18.29$\pm$0.06 mag, $J$=17.60$\pm$0.17 mag,
$H$=17.24$\pm$0.14 mag, and $K_{\rm s}$=17.57$\pm$0.18 mag. Using
the LePHARE simulation program\cite{Arn14} and the NextGen model
atmosphere grid\cite{Hau99}, and leaving the foreground reddening
towards the source as a free parameter, we found that the best
fitting SED ($\chi^2$=2.0; Extended Data Fig.\,1) is that of a
star with an effective temperature of $T_{\rm eff}\approx$5200 K
and a colour excess of $E(B-V)\approx$0.9 mag.

\subsection{VLT/FORS2 spectroscopy of [GV2003]\,N.}
%
[GV2003]\,N was observed once each night on 2015 April 14, 21, May
13 and 16 with the VLT/FORS2 instrument. The observations were
conducted using a long slit with 1.0 arcsec width centred on the
target. We used the 1200R grism in order to obtain a continuous
coverage of the $\approx$5870--7370\,\AA \, wavelength range with
an average spectral resolution, measured using the emission lines
of the wavelength calibration lamps, of
$\Delta$$\lambda$/$\lambda$$\approx$2900. Each exposure was
2700\,s long. For the parameter determination and abundance
analysis, the four spectra have been co-added yielding a S/N per
pixel of $\approx$93 at $\approx$6800\,\AA. The science images
were reduced using a set of ten bias and ten flat-field images
collected on the morning following each observing night. Each
spectrum was extracted using an aperture of 12 pixels and by
applying background subtraction. The spectra were wavelength
calibrated using a wavelength calibration lamp obtained on the
morning following each observing night. We further refined the
absolute wavelength calibration using the sky Na\,{\sc i}
$\lambda\lambda$5890, 5896 emission lines, with a typical
correction of $\approx$$2\,\kms$.

In order to derive an accurate RV from each spectrum of the star,
we co-added the spectral lines using the Least-Squares
Deconvolution technique (LSD), which is effective also with FORS
spectra obtained with the 1200 grisms\cite{Bag12}. The LSD
technique\cite{Koc10} combines line profiles centred at the
position of the individual lines given in the line mask and scaled
according to the line strength. The resulting average profiles
were obtained by combining about 300 lines, yielding a strong
increase in S/N, therefore improving the precision of the RV
measurements. We prepared the line mask used by the LSD code,
adopting the stellar temperature derived from the GROND SED, a
surface gravity of $10^{4.5}$ typical of main-sequence solar-like
stars, and solar abundances\cite{Asp09}. We note that the FORS2
spectrum does not cover strong enough lines of two ionization
stages of the same element to spectroscopically infer the surface
gravity. We extracted the line parameters from the Vienna Atomic
Line Database (VALD; ref.\cite{Pis95}) using all lines stronger
than 20\% of the continuum, avoiding hydrogen lines and lines in
spectral regions affected by the presence of telluric and nebular
features.

The RV values, obtained by fitting a Gaussian to each LSD profile,
are listed in Table\,\ref{tab:RV}. The results show the presence
of clear RV variations, indicative of binarity, with a period of
the order of about a month, or less. The large RV variation
derived from the two observations conducted in May, in contrast to
that obtained from the April data on a similar time-base, is
suggestive of an eccentric orbit, which is typical of young
post-SN binaries\cite{Hil83}.

\subsection{Spectral analysis.} \label{sec:spec}
Because of the moderate resolution and hence strong line blending,
the spectrum of [GV2003]\,N could be analyzed only through
spectral synthesis. We calculated synthetic spectra with {\sc
synth3}\cite{Koc07} on the basis of line lists extracted from the
VALD database and of model atmospheres calculated with the {\sc
LLmodels} stellar model atmosphere code\cite{Shu04}. For all
calculations we assumed Local Thermodynamical Equilibrium (LTE),
plane-parallel geometry, and a microturbulence velocity of $1.0 \,
\kms$, typical of solar-like stars (e.g. ref.\cite{Bru10}).

The effective temperature of solar-like stars can be best
estimated using the wings of hydrogen lines, H$\alpha$ in
particular\cite{Fuh93}, which is fully covered by the FORS2
spectra. Unfortunately, the intensity of the nebular H$\alpha$
emission varies strongly in the region around [GV2003]\,N,
preventing a reliable background subtraction at the wavelengths
covered by the H$\alpha$ line. We therefore derived $T_{\rm eff}$
from the analysis of the Fe\,{\sc i} lines. We selected a set of
15 strong and weakly blended Fe\,{\sc i} lines from which we
derived the iron abundance\cite{Fos07}. By imposing no correlation
between line abundance and excitation energy and taking into
account the results obtained from the SED, we estimated an
effective temperature of 5100$\pm$200\,K. The broadening of the
spectral lines is dominated by the instrumental resolution; this
sets an upper limit on the stellar projected rotational velocity
$v\sin i$ of about $80 \, \kms$. Because of the lack of
spectroscopic $\log g$ indicators, we adopted the stellar surface
gravity obtained from comparing the position of the star in the
Hertzsprung–Russell diagram with evolutionary tracks, as described
in the main text.

On the basis of the derived stellar parameters, we derive the
abundance of Si, Ca, Ti, V, Cr, Mn, Fe, Co, Ni, and Ba
(Fig\,\ref{fig:abund}). Because of the rather large
overabundances, we followed the same procedure typically adopted
for a self-consistent analysis of magnetic chemically peculiar
stars, hence iteratively re-calculating a new grid of model
atmosphere, atmospheric parameters, and abundances each time the
abundances changed significantly from the previous
iteration\cite{Shu09}. This was possible because the model
atmosphere code allows the use of individualized abundance
patterns.

After convergence, we derived the statistical abundance
uncertainty for each element, i.e., the standard deviation from
the average abundance given a set of two or more lines, and the
maximum systematic uncertainty, which takes into account the error
bars on the atmospheric parameters: 200\,K for $T_{\rm eff}$,
0.1\,dex for $\log g$, and $0.5 \, \kms$ for the microturbulence
velocity. The statistical uncertainty accounts for uncertainties
in the placement of the continuum and in the atomic data. For the
elements for which we derived the abundance from one line, we
assumed that the statistical uncertainty is equal to that of Fe.
The true uncertainty lies in between the statistical and maximum
systematic uncertainties\cite{Fos09}.

The spectrum does not show the presence of the Li\,{\sc i}
$\lambda$6708\,\AA \, line; we could therefore derive an upper
limit on the Li abundance. The Mn and Ba abundances were derived
taking into account hyperfine structure. The resulting abundances
for all considered elements are listed in Extended Data Table\,2.

\end{methods}

\section*{Bibliography}
\vspace{1cm}

\newpage

\noindent {\bf \large Supplementary Information}

\noindent {\bf 1. \, RCW\,86 as the result of a SN explosion near
the edge of a wind-blown bubble}

RCW\,86 (also G315.4--2.30, MSH\,14--6{\it 3}) is a young Galactic
shell-like SNR with an angular diameter of about 40$^\prime$,
which -- for a distance to the remnant of 2.3 kpc
(ref.\cite{Sol03}) -- corresponds to $\approx$26\,pc. The almost
complete shell of RCW\,86, visible in X-ray\cite{Bor01},
optical\cite{Smi97}, radio\cite{Whi96} and infrared\cite{Wil11}
wavelengths, has a bright peculiar protrusion to the south-west,
what gives the SNR a pyriform appearance (see fig.\,6 in
ref.\cite{Smi97}). At optical wavelengths, the protrusion appears
as a ragged arc-like nebula\cite{Rod60, Ros96} (Fig.\,1) with a
radius of about $2^\prime$ (1.1 pc). Spectroscopic observations of
the arc indicate that the SN blast wave expands with a velocity of
$\sim 100 \, \kms$ at this location. Deep imaging of RCW\,86
revealed that almost the whole periphery of the SNR is encompassed
by Balmer-dominated filaments\cite{Smi97}. Spectra of these
filaments indicate shock velocities of $\approx$$500-900 \, \kms$
(ref.\cite{Lon90, Sol03}). Similar velocities of the SN blast wave
were also inferred from observations of thermal X-ray emission in
RCW\,86\cite{Bor01}, while the detection of nonthermal X-ray
emission in this SNR\cite{Bam00, Bor01, Rho02} implies the
existence of shocks with even higher speeds. Recent proper motion
measurement of a nonthermal X-ray filament in the north-east rim
of RCW\,86 showed\cite{Yam16} that the blast wave expands there
with a speed of $\approx$$3000 \, \kms$.

The striking asymmetry in expansion velocity of the SN blast wave
was interpreted as an indication that the SN explosion has
occurred within a low-density bubble created by the wind of the
progenitor star\cite{Vin97}, and that the actual SN explosion
centre is closer to the south-west edge of the bubble because the
ambient medium in this direction is denser than in the opposite
(north-east) one\cite{Vin00}. Correspondingly, it was suggested
that the SN blast wave has already hit the nearest (south-west)
edge of the bubble and slowed down its velocity to $\approx$$600
\, \kms$, while in the opposite direction it still freely expands
within the bubble. This scenario is widely accepted in subsequent
papers\cite{Wil11, Bro14} on RCW\,86, in particular, to explain
the young kinematic age of the SN blast wave implied by the
possible association of this SNR with the historical supernova of
A.D. 185 (SN\,185)\cite{Cla75}.

Gvaramadze \& Vikhlinin\cite{Gva03} supplemented the cavity
explosion scenario by Vink et al.\cite{Vin97} by suggesting that
the cavity was created by a moving star and that the SN explosion
in RCW\,86 has occurred near the edge of the cavity -- in the
centre of the south-west protrusion, which is similar to the model
put forward by Wang et al.\cite{Wan93} to explain the origin of
large-scale structures (in particular the Napoleon's Hat) around
the SN\,1987A. An interesting consequence of this interpretation
is that after the blast wave will have completely overrun the
south-west protrusion, RCW\,86 will assume a two-shell form, with
a newly-formed shell attached to the already existing one. This
points to the possibility that some of the known two-shell SNRs
could originate from off-centred cavity SN explosions\cite{Gva07}.

Including our results, we arrive at the following scenario. The SN
progenitor was a $\sim$$10 \, \msun$ star in a tight binary system
with a low-mass companion and the SN explosion was of Type\,Ib.
This means that before the SN explosion the binary system has lost
a significant fraction of its initial mass, $M_{\rm i}$, through a
common-envelope ejection. The ejected mass is expected to be about
$5 \, \msun$, since (single) $10 \, \msun$ stars lose only about
10\% of $M_{\rm i}$ during their lives, and the pre-SN mass of the
exploding star is constrained to $\leq$$4 \, \msun$. After the
common-envelope ejection, the stripped star could still be
surrounded by a thin hydrogen envelope\cite{Yoo10}, which is blown
off later on through a fast line-driven wind. This wind would
sweep-up the material of the common-envelope ejecta in a dense
shell, which we now observe as an arc-like nebula in the
south-west corner of RCW\,86.

From the expected ejection velocity of the common envelope of
several tens of $\kms$ and given the radius of the optical arc of
$\approx$$1.1$ pc, one finds that the envelope ejection event
occurred a few $10^4$ yr before the SN explosion. This time-scale
is similar to the duration of the shell helium burning phase of
$\sim$$10 \, \msun$ stars\cite{Jon13}, which implies that a
case\,C Roche-lobe overflow initiated the common envelope
evolution. Using the number density of the local interstellar
medium of $\sim$$0.5 \, {\rm cm}^{-3}$ (ref.\cite{Wil11}), one
finds that the common-envelope ejecta so far swept-up only
$\sim$$0.1 \, \msun$ of the ambient gas, which in turn implies
that the arc should be composed mostly of the material lost by the
binary system, and that it should therefore be overabundant in
nitrogen by a factor of three or so\cite{Bro11}. Optical
spectroscopy of filaments at several positions in the optical arc
revealed\cite{Rui81} a factor of two overabundance of nitrogen
with respect to the value typical of old Galactic SNRs. Although
this overabundance is slightly lower than what is expected from
our scenario, we note that the observed spectra could be a
superposition of spectra produced in regions with different
physical conditions (e.g. because of the presence of shocks with
different velocities along the same line-of-sight), which makes it
difficult to determine the nitrogen and other chemical abundances
with standard methods (cf. ref.\cite{Rui81}).

\noindent {\bf 2. \, Chemical abundances in LMXBs}

Heavy element enrichment due to an exploding binary companion have
been measured in a handful of stars\cite{Isr99, Gon05, Gon11,
Sua15}, all in active X-ray binaries, where the X-rays produced by
mass transfer from a low mass main sequence star to a compact
companion prove the physical connection of both components. Except
for one object (Nova Scorpius), the pollution of the low mass star
with SN products is rather mild, i.e., some metal abundances are
typically enhanced by a factor of two, with an oxygen enhancement
of ten in Nova Scorpius as the record holder\cite{Isr99}.
Remarkably, the Ca overabundances of all objects so far are
smaller than a factor of 1.6, including that in Nova Scorpius (see
table\,5 in ref.\cite{Gon11}). This is to be contrasted with the
Ca enhancement of a factor of 6 in [GV2003]\,N.

More importantly, in the known post-SN binaries Si is more
overabundant than Ca\cite{Gon11}. This is consistent with
studies\cite{Woo95, Chi13} of explosive yields from (single) stars
with $M_{\rm i}\geq$$11 \, \msun$, which show that Si is generally
produced in larger amounts than Ca. While, in principle, selective
accretion could still lead to the observed abundance pattern in
[GV2003]\,N, we consider this unlikely since the velocities of the
Ca- and Si-rich layers in the ejecta of a normal core-collapse SN
are similar. We suggest therefore that the elevated overabundance
of Ca with respect to Si is intrinsic to the SN ejecta, which
points to the possibility that RCW\,86 is the result of a Ca-rich
SN explosion (see next section).

\noindent {\bf 3. \, Calcium-rich SNe}

Ca-rich SNe are a recently defined class of faint and rapidly
evolving SNe whose observational properties appear challenging to
explain. While at maximum light, the optical spectra resemble
those of Type\,Ib SNe, thus showing lines of helium but not of
hydrogen\cite{Kas12}, these objects quickly evolve to the
optically thin nebular phase (at $\sim$2 months), where their
optical spectra are dominated by [Ca\,{\sc ii}], with lines from
oxygen or other intermediate mass elements being weak or absent.

The question which progenitors can produce Ca-rich SNe is not
settled. These SNe appear to occur in early type galaxies and many
of them are found at large (up to 150 kpc) galactocentric
distances\cite{Lym14, Fol15}. This points towards relatively large
progenitor lifetimes (from tens to hundreds of Myr) and/or very
high ($\sim$$1000 \, \kms$) space velocities\cite{Fol15}, which in
turn implies low-mass SN progenitor stars. However, a fraction of
the Ca-rich SNe are found within their host galaxies\cite{Lym16b}
and many of the host galaxies show signs of galaxy-galaxy
interaction or merger\cite{Fol15, Kaw10}. The latter implies that
the host galaxies of the Ca-rich SNe might be sites of recent
(massive) star formation and that some of these SNe might
originate from massive stars. Correspondingly, thermonuclear
events from white dwarf progenitors\cite{She09, Per10, Wal11,
Sel15} and core-collapse events produced from stripped lowest mass
($M_{\rm i}\sim$$10 \, \msun$) massive stars\cite{Kaw10, Mor10}
have been proposed to explain the Ca-rich SNe. In both cases,
explosive helium burning is expected to occur. Whereas SN
simulations of stripped stars are available only for stars with
rather large $M_{\rm i}$ (e.g. ref.\cite{Des16} and references
therein), Waldman et al.\cite{Wal11} found in simulations of
helium shell detonations on CO white dwarfs that Ca may be the
most abundant product in such events, together with large amounts
of unburnt helium.

In parameterized explosive helium burning conditions, Hashimoto et
al.\cite{Has83} find Ca to be the main product for a peak
temperature of $\approx$$1.3\times10^9$\,K. Assuming the core
radius of a white dwarf of $r=0.01 \, R_{\odot}$ and adopting the
SN post-shock temperature $T_{\rm ps}=[3E_{\rm SN}/(4\pi r^3
a)]^{1/4}$ from ref.\cite{Wea80}, with $E_{\rm SN}$ being the SN
kinetic energy and $a=7.57\times10^{-15} \, {\rm erg} \, {\rm
cm}^{-3} \, {\rm K}^{-4}$ being the radiation constant, one finds
that $T_{\rm ps}$ would reach the above value if $E_{\rm
SN}=3\times10^{49}$\,erg. Weaver \& Woosley\cite{Wea80} argue for
a shock-induced production of Ca at somewhat higher temperature of
$T_{\rm ps}$$\approx$$3.5\times10^9$\,K, which would require
$E_{\rm SN}$$\sim$$10^{51}$\,erg. In the lowest mass stripped
core-collapse SN progenitors, the mass of the mantle of
intermediate mass elements surrounding the innermost strongly
electron-degenerate core is extremely small\cite{Wan09, Jon13}.
Thus the massive helium shell --- which is strongly diminished in
single stars due to the second dredge-up\cite{Pod04}, but which
remains intact in stripped stars --- is located at a very small
radius and can therefore be exposed to very high temperatures when
the SN shock wave passes through. It thus appears possible, but
remains to be shown with detailed models, that SNe from stripped
stars of relatively low $M_{\rm i}$ produce fast and faint
transients with Ca- and helium-rich ejecta. This possibility is
supported by the finding that the light curves produced by
stripped electron-capture SNe are consistent with those of the
Ca-rich SNe\cite{Mor16}.

\noindent {\bf 4. \, Simulated effects of a SN explosion in a
tight binary}

We performed Monte Carlo calculations of 10\,000 asymmetric SN
explosions to simulate the resulting kinematics on the surviving
binary system containing a NS and a $0.9 \, \msun$ G-dwarf, based
on the equations of Tauris \& Takens\cite{Tau98}. The impact of
the SN shell is modelled according to the three-dimensional
hydrodynamical simulations results of Liu et al.\cite{Liu15}, by
applying their power-law fit (see their fig.\,6) for a $0.9 \,
\msun$ main-sequence companion star which experiences shell impact
from a nearby, stripped exploding core of $3.0 \, \msun$.

The mass of the exploding star and the pre-SN orbital period are
randomly drawn from the intervals $2.9-3.7 \, \msun$ and
$0.28-1.0$\,d. The chosen mass interval corresponds to $M_{\rm i}$
in the range $10-13 \, \msun$ (see main text). A minimum pre-SN
orbital period of 0.28\,d was adopted to prevent the $0.9 \,
\msun$ star from filling its Roche lobe, which would most likely
lead to a merger event. The upper limit of the orbital period is
given by the strong pollution of the companion star with heavy
elements from the SN ejecta, which means that the pre-SN orbit
must have been tight. The momentum kicks are randomly drawn from
the range of $0-500 \, \kms$, assuming an isotropic distribution
of kick directions. We assume a NS with a mass of $1.4 \, \msun$
is formed, releasing a gravitational binding energy of $0.16 \,
\msun \, c^2$ (ref.\cite{Lat89}). Since we estimate a post-SN
orbital period in the range of $10-30$\,d (Methods) we consider
only cases where this is achieved.

Extended Data Fig.\,2 (left panel) shows that the most likely
obtained systemic velocities are of the order of $80 \, \kms$, in
agreement with observational constraints based on the location of
the NS relative to the centre of the optical arc in the south-west
of RCW\,86. We found a minimum orbital eccentricity of 0.7. Liu et
al.\cite{Liu15} obtained the mass of accreted SN ejecta by the
companion star as function of $E_{\rm SN}$ and the pre-SN binary
separation. We parameterized their results, and applied them to
our calculations. Extended Data Fig.\,2 (right panel) shows that
using the canonical value of $E_{\rm SN}=10^{51} \, {\rm erg}$
results in very small accreted amounts of SN ejecta (blue
histogram, left), but a five times smaller SN energy yields values
(red histogram, right) which are well compatible with the
enrichments seen in [GV2003]\,N (see main text). A smaller than
canonical supernova energy is also consistent with our initial
mass estimate of the SN progenitor of about $10\,M_{\odot}$ (cf.
ref.\cite{Kaw10, Tau15}).

\noindent {\bf \large References} \vspace{1cm}


\pagebreak

\begin{figure}
\caption{{\bf Extended Data Figure 1\quad Observed spectral energy
distribution of [GV2003]\,N.} Data points (filled red circles)
show the GROND $g'r'i'z'JHK_{\rm K}$ magnitudes. The horizontal
error bars represent the full-width at half maximum of the GROND
filters. The quadratic sum of the statistical uncertainties and
the systematic uncertainties of the brightness measurement are
shown as vertical error bars (in most cases they are within the
size of the data points). The best fitting spectral energy
distribution (solid line) is that of a star with an effective
temperature of $\approx$5200 K and the colour excess of
$E(B-V)$=0.9 mag.} \label{fig:SED}
\begin{center}
\includegraphics[width=16cm,clip]{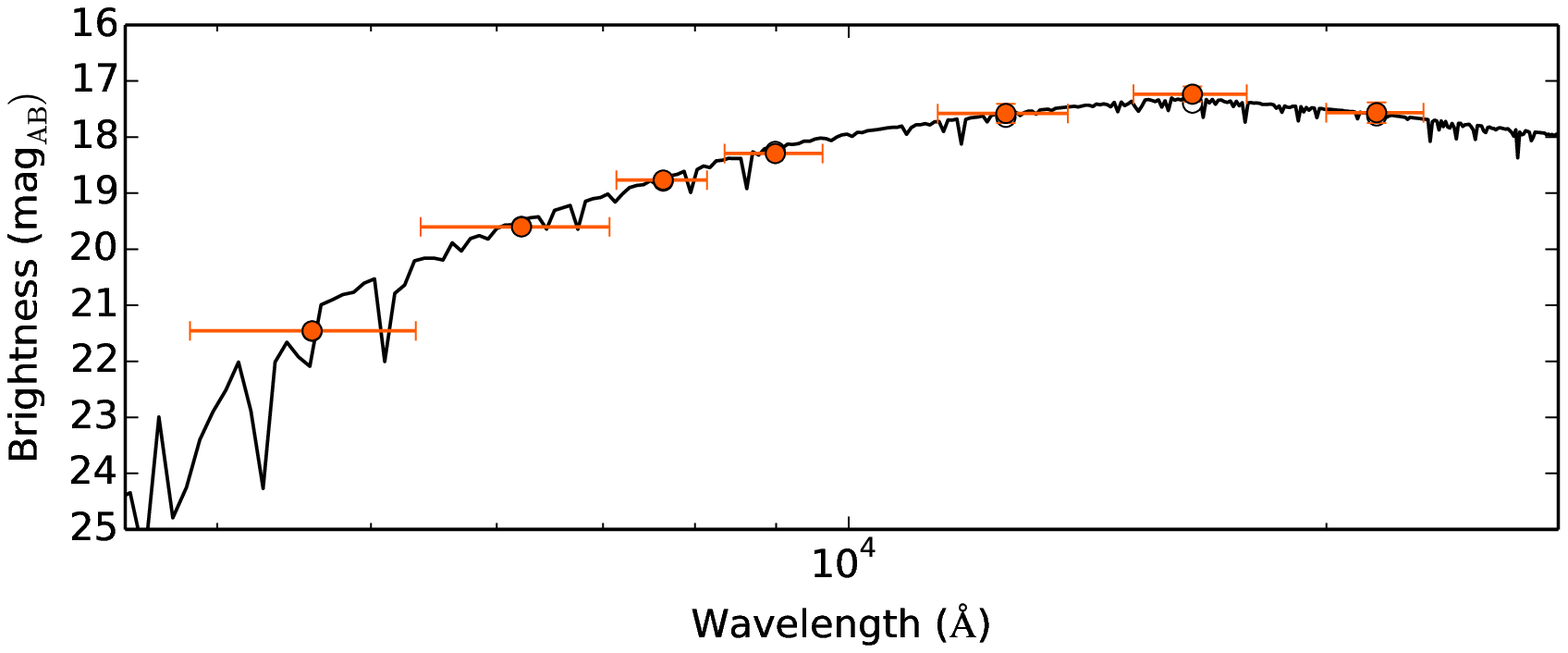}
\end{center}
\end{figure}

\pagebreak

\begin{figure}
\caption{{\bf Extended Data Figure 2\quad Results of simulations
of 10,000 asymmetric SNe in tight binary systems}. The binary
systems were assumed to be composed of an exploding $2.9-3.7 \,
\msun$ core and a $0.9\msun$ main sequence star with an orbital
period between 0.28 and 1.0\,d. Kick velocities were chosen
randomly in the range from 0 to $500 \, \kms$. The left panel
shows the resulting systemic velocities of the post-SN binaries.
The right panel gives the amount of SN ejecta accreted by the main
sequence star, calculated for two values of the SN kinetic energy:
$E_{SN}=10^{51}$\,egr (blue histogram, left) and
$E_{SN}=2\times10^{50}$\,egr (red histogram, right).
\label{fig:kicks}}
\begin{center}
\includegraphics[width=0.4\columnwidth,angle=270]{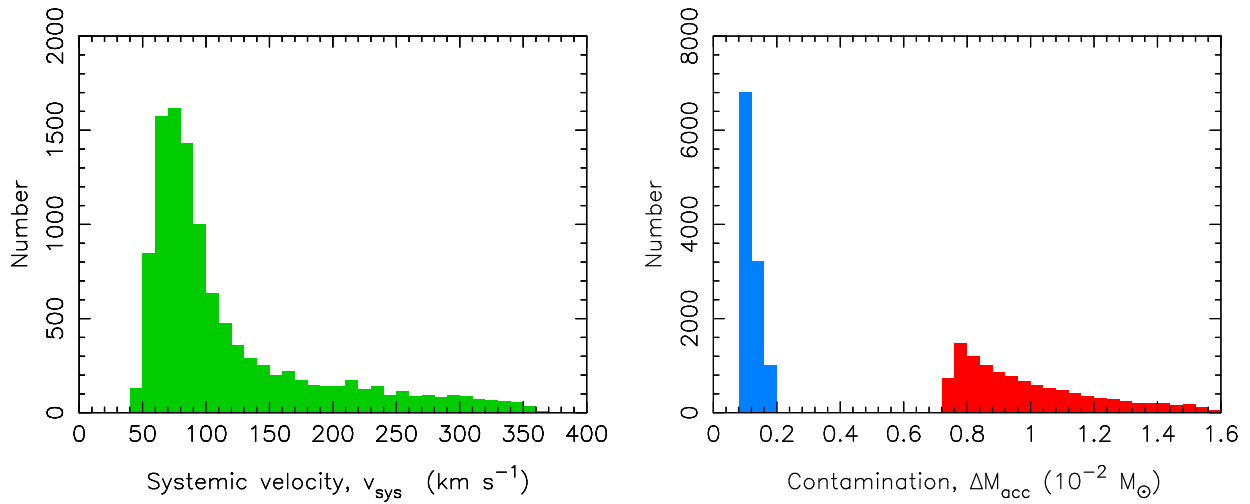}
\end{center}
\end{figure}

\newpage

\begin{table}
  \caption{\bf Radial velocity changes with time in the spectrum of
  [GV2003]\,N}
  \label{tab:RV}
  \renewcommand{\footnoterule}{}
  \begin{center}
   \begin{tabular}{lll}
      \hline
Date & Radial & Heliocentric \\
& velocity$^*$ & correction \\
& ($\kms$) & ($\kms$) \\
\hline
2015 April 14 & $-71$$\pm$2 & 13.0 \\
2015 April 21 & $-74$$\pm$4 & 10.9 \\
2015 May 13 & $-44$$\pm$3 & 3.3 \\
2015 May 16 & $-62$$\pm$3 & 2.4 \\
\hline
    \end{tabular}
    \end{center}
$^*$ Corrected for the heliocentric motion of the Earth.
    \end{table}

\newpage

\begin{table}
\caption{{\bf Extended Data Table 1\quad Surface element
abundances of [GV2003]\,N} \label{tab:abun}}
\renewcommand{\footnoterule}{}
\begin{center}
\begin{tabular}{lll}
\hline
Element & $X/H$ & $ M_{\rm acc}/(10^{-5} \, \msun)$ \\
\hline
Li  &  $<$0.29$^*$               &  --   \\
Si  &  0.24$\pm$0.10(0.13)$^\dagger$    &  8 \\
Ca  &  0.81$\pm$0.05(0.24)    &  7 \\
Ti  &  0.50$\pm$0.18(0.46)    &  0.09 \\
V   &  0.67$\pm$0.01(0.37)    &  0.028 \\
Cr  &  0.47$\pm$0.10(0.23)    &  0.6 \\
Mn  &  0.62$\pm$0.18(0.27)    &  0.8 \\
Fe  &  0.22$\pm$0.10(0.19)    &  15 \\
Co  &  0.38$\pm$0.04(0.23)    &  0.09 \\
Ni  &  0.45$\pm$0.33(0.35)    &  1.8 \\
Ba  &  $-$0.04$\pm$0.10(0.43) &  0   \\
\hline
\end{tabular}
\end{center}
Abundances for elements are $X/{\rm H}=\log [N(X)/N({\rm H})]_{\rm
star}-\log [N(X)/N({\rm H})]_{\odot}$, where $N(X)$ is number
density of atoms. In the third column we provide the implied
amount of accreted mass per element assuming dilution in a
convective envelope of $0.2 \, \msun$. \\
$^*$ For the Li abundance only an upper limit is derived (see
Methods).\\
$^\dagger$ The uncertainties quoted for the abundances are
statistical and maximum systematic (in brackets). The true
uncertainties lie in between. \\
\end{table}

\end{document}